\documentclass[
 aip, apl,
 amsmath,amssymb,
 reprint,
]{revtex4-1}

\usepackage{graphicx}
\usepackage{dcolumn}
\usepackage{bm}
\usepackage[mathlines]{lineno}
\usepackage{xcolor}
\usepackage[utf8]{inputenc}
\usepackage[T1]{fontenc}
\usepackage{mathptmx}

\begin{document}

\preprint{AIP/123-QED}

\title[TACIT APL]{Indium-Bond-And-Stop-Etch (IBASE) Technique for Dual-side Processing of Thin High-mobility GaAs/AlGaAs Epitaxial Layers}

\author{C. Yoo}
\affiliation{Physics Department and Institute for Terahertz Science and Technology, University of California, Santa Barbara, California 93106 USA}

\author{K. W. West}
\affiliation{Department of Electrical Engineering, Princeton University, Princeton, New Jersey 08544 USA}
 
\author{L. N. Pfeiffer}
\affiliation{Department of Electrical Engineering, Princeton University, Princeton, New Jersey 08544 USA}
 
\author{C. A. Curwen}
\affiliation{Jet Propulsion Laboratory, California Institute of Technology, Pasadena, California 91109 USA}

\author{J. H. Kawamura}
\affiliation{Jet Propulsion Laboratory, California Institute of Technology, Pasadena, California 91109 USA}

\author{B. S. Karasik}
\affiliation{Jet Propulsion Laboratory, California Institute of Technology, Pasadena, California 91109 USA}
 
\author{M. S. Sherwin\textsuperscript{*}}
 \email{sherwin@ucsb.edu}
\affiliation{Physics Department and Institute for Terahertz Science and Technology, University of California, Santa Barbara, California 93106 USA}
 
\date{\today}

\begin{abstract}
We present a reliable flip-chip technique for dual-side processing of thin (<1 $\mu$m) high-mobility GaAs/AlGaAs epitaxial layers. The technique allows the fabrication of small (micron-scale with standard UV photolithography) patterned back gates and dual-gate structures on the thin GaAs/AlGaAs films with good alignment accuracy using only frontside alignment. The technique preserves the high-mobility (>10$^{6}$ cm$^{2}$/V-s at 2 K) and most (>95\%) of the charge density of the 2-dimensional electron gas (2DEG) systems, and allows linear control of the charge density with small (< 1 V) electrostatic gate bias. Our technique is motivated by a novel THz quantum-well detector based on intersubband transitions in a \textit{single}, wide GaAs/AlGaAs quantum well, in which a symmetric, well-aligned dual-gate structure (with a typical gate dimension of $\sim 5 \mu $m $ \times 5 \mu$m) is required for accurate and precise tuning of the THz detection frequency. Using our Indium-Bond-And-Stop-Etch (IBASE) technique, we realize such dual-gate structure on 660-nm thick GaAs/AlGaAs epitaxial layers that contain a modulation-doped, 40-nm wide, single square quantum well. By independently controlling the charge density and the DC electric field set between the gates, we demonstrate robust tuning of the intersubband absorption behavior of the 40-nm quantum well near 3.44 THz at 30 K. 
\end{abstract}

\maketitle

Dual-side processing (or backside processing) is important for defining patterned back gates and dual-gate structures required in various high-mobility GaAs/AlGaAs heterostructure devices.\cite{Eisenstein1990, Croxall2013, Sherwin1997} Of particular interest in this Letter is a recently demonstrated tunable antenna-coupled intersubband terahertz (TACIT) detector that uses intersubband transitions in a \textit{single}, wide GaAs/AlGaAs quantum well (QW) for efficient absorption of THz radiation.\cite{Sherwin1997, Sherwin2002, Yoo2020} In the TACIT detector, the detection frequency is determined by the intersubband absorption of a 2-dimensional electron gas (2DEG) .\cite{Sherwin1997, Sherwin2002, Yoo2020} For the relatively wide quantum wells appropriate for TACIT detectors, the intersubband absorption frequency can be significantly tuned from the single-electron value predicted for flat band conditions by applying a DC electric field in the growth direction (DC Stark effect), and by varying the charge density (many-body effects).\cite{Helm1999} To take advantage of this tunability, both the charge density and DC electric field must be independently controlled, requiring a symmetric, well-aligned dual-gate structure on the QW structure with a typical thickness less than 1 $\mu$m. In addition, because the read-out mechanism of the TACIT detector depends on the bolometric response associated with temperature-dependent phonon scattering in a high-mobility 2DEG, it is important to fabricate the dual-gate structure without sacrificing its high mobility or charge density.

Typically, such dual-gate structures can be defined using regrowth \cite{Linfield1993, Evans1994, Brown1994,Berl2016} or flip-chip \cite{Eisenstein1990,Evans1994, Blount1996,Croxall2008,Gupta2012} techniques. In regrowth techniques, highly-doped conductive GaAs layers are grown first with molecular beam epitaxy (MBE) prior to the regrowth of active QW layers. These conductive layers are selectively damaged before the regrowth to form patterned back gates, either in-situ via focused ion beam\cite{Linfield1993, Brown1994} or ex-situ via selective wet etch\cite{Evans1994} or oxygen implantation.\cite{Berl2016} While these regrowth techniques can be compatible with extremely high-mobility devices,\cite{Berl2016} they require either expensive focused-ion beam equipment (for the in-situ method) or relatively complex processing and cleaning steps (for the ex-situ methods). Additionally, because these regrowth techniques pre-define the patterns for the back gates at the wafer level, they are less robust to design variations after the full growth of the QW wafer. Lastly, accurate alignment can be challenging due to the typical lack of alignment features after the regrowth. 

Another widely used method are the flip-chip techniques.\cite{Eisenstein1990, Blount1996, Croxall2008, Gupta2012} In these techniques, the front side of a QW sample or "die" is processed first and then flip-chip bonded to a host substrate for backside processing. During the backside processing, the sacrificial back side of the die is removed, and standard lithographic patterning and metallization steps are used to define patterned back gates directly on the thinned back side. In the most recent versions of these techniques, represented by the well-known Epoxy-Bond-And-Stop-Etch (EBASE) technique\cite{Blount1996} and its variants,\cite{Croxall2008,Gupta2012} the QW die is thinned down using etch-stop process to extremely thin (<1 $\mu$m) high-mobility GaAs/AlGaAs layers. In these techniques, however, significant backside processing steps (such as via formation\cite{Blount1996} or application of silver paste\cite{Gupta2012}) are required to electrically access frontside contacts that are buried after flip-chip bonding. Therefore, achieving reliable and robust electrical contacts without sacrificing device yield remains challenging.

\begin{figure*}[ht!]
\includegraphics[width=1\textwidth]{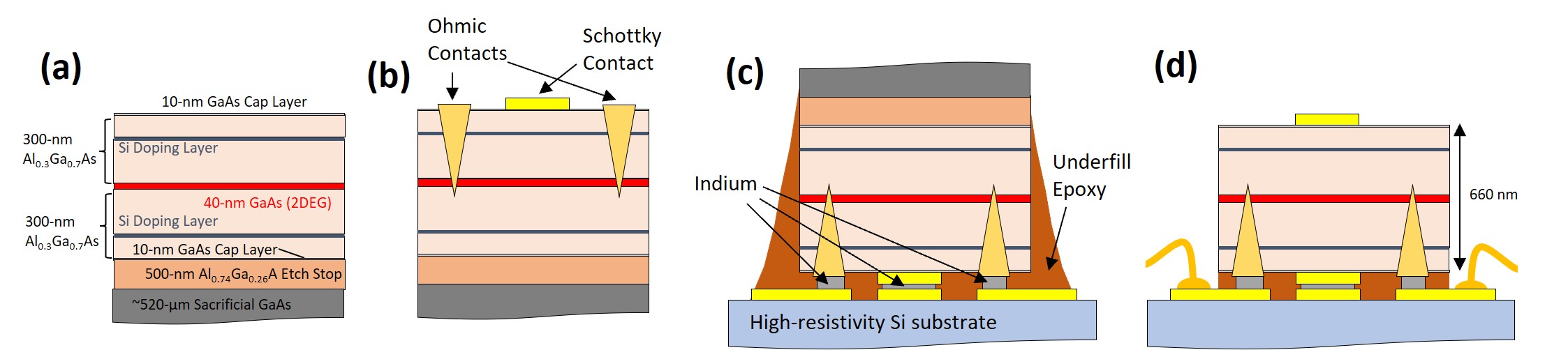}
\caption{Flow diagrams for the Indium-Bond-And-Stop-Etch (IBASE) process. (a) Simplified schematic for our QW sample (not to scale). (b) Frontside fabrication (c) Flip-chip bonding and underfilling. (d) Backside fabrication.}
\label{fig:fig1}
\end{figure*}

In this Letter, we demonstrate an alternative, robust flip-chip technique that allows the dual-side processing of thin (<1 $\mu$m) high-mobility GaAs/AlGaAs epitaxial layers without requiring complex backside processing steps for reliable electrical contacts. Our Indium-Bond-And-Stop-Etch (IBASE) technique allows good alignment accuracy ($< 0.5 \mu$m) with standard UV contact aligners without special IR backside alignment, and offers reliable electrical access to frontside contacts using indium bonding. Standard Hall measurements on the sample devices fabricated with the IBASE process show that our gating technique is compatible with high-mobility (>10$^{6}$ cm$^{2}$/V-s at 2 K) GaAs/AlGaAs 2DEG systems and that linear control of charge density with a small (< 1 V) DC gate bias is possible. Finally, as a practical application of the IBASE technique, we fabricate a dual-gate structure for a prototype TACIT detector\cite{Yoo2020} and demonstrate robust tuning of the THz intersubband absorption behavior of the device by independently controlling the charge density and the DC electric field set between the gates.

Figure \ref{fig:fig1} illustrates the IBASE process for the fabrication of a dual-gate structure on our QW sample. The process begins with the MBE growth of the QW epitaxial layers on a sacrificial GaAs substrate (Fig.\ref{fig:fig1}a). \footnote{While our gating technique can be applied to a variety of QW heterostructures including double-QWs and superlattices, we used a single, square QW for the fabrication of the devices presented in this Letter.} As in other flip-chip techniques,\cite{Blount1996,Croxall2008,Gupta2012} an AlGaAs layer with a higher ($\sim$74 \%) Al content is used as an etch-stop layer for complete removal of the sacrificial backside GaAs.\cite{Blount1996} Our sample had a 40-nm wide, single, square QW with the center of the 2DEG layer placed 330 nm below the top surface. The 2DEG was remotely doped with Si layers placed 220 nm away from the center of the 2DEG. For the devices described in this Letter, we used two versions of the QW structures, labeled as QW1 and QW2, with two different Si doping levels. For QW1, the charge density $n_{s}$ and mobility $\mu$ (measured with Hall measurements in the dark at 2 K) were $2.5 \times 10^{11}$ cm\textsuperscript{-2} and $7.0\times 10^{6}$ cm\textsuperscript{2}V\textsuperscript{-1}s\textsuperscript{-1}, respectively. For QW2, we had $n_{s} = 2.1 \times 10^{11}$ cm\textsuperscript{-2} and $\mu = 7.3\times 10^{6}$ cm\textsuperscript{2}V\textsuperscript{-1}s\textsuperscript{-1}. Further details on the growth of our QW structures are provided in the supplementary material.

Following the MBE growth, the front side of the QW sample is first processed (Fig.\ref{fig:fig1}b). During the frontside processing, a 2DEG mesa is defined, and frontside (Ohmic or Schottky) contacts are fabricated using standard lithographic and metallization steps. After the fabrication of the frontside contacts, the QW sample, which may contain several identical devices at this point, is diced into smaller (e.g., 3 mm x 3 mm) dies for individual flip-chip bonding. 

Once the frontside fabrication is complete, a host substrate is prepared for flip-chip bonding (Fig. \ref{fig:fig1}c). The host substrate can be either intrinsic GaAs or Si, and contains metal electrodes that match the frontside contacts of the QW die and indium bumps required for flip-chip bonding. Standard lithographic patterning and metallization steps are used to define both the electrodes and the indium bumps. The typical thickness for the indium bumps is $\sim$ 2 $\mu$m to avoid any mechanical damage on the frontside contacts (typically less than 1-$\mu$m thick) during the flip-chip bonding. After both the electrodes and the indium bumps are defined, the substrate piece, which may also have multiple identical host substrates, is diced into smaller pieces, with the individual substrate size slightly larger (e.g., 4 mm x 4 mm for our case) than the QW die. The size of the substrate is intentionally chosen to be larger to provide excess area on the substrate for wire-bonding pads to be defined.

After the host substrate is prepared, the QW die is flip-chip bonded to the host substrate (Fig. \ref{fig:fig1}c). A flip-chip bonder (e.g. Finetech lambda) is used to align and place the QW die on the substrate. A gentle force ($\sim$ 3 N) is applied on the back side of the QW die while heating up both the QW die and the Si substrate to 180 $^\circ$C to reflow indium for bonding. For better bonding quality, the native oxide on the indium is removed with a diluted HCl solution prior to the bonding. The indium bonds establish direct electrical connection between the frontside contacts of the QW die and the matching electrodes on the substrate. After the bonding, the rest of the area on the bonded surfaces is filled with underfill epoxy to prevent mechanical and chemical damage throughout the rest of the IBASE process. The choice of underfill epoxy and its curing condition is critical for the success of the rest of the IBASE process, as any unwanted air bubbles trapped during underfilling can cause significant surface damage during selective backside etches. We found that underfill epoxy with low viscosity (EPO-TEK 353ND) cured at low temperature for prolonged duration (at 80 $^\circ$C on a hot plate for an hour) gives the best results that minimize air bubbles and provide good mechanical and chemical protection. 

Once the underfill epoxy is cured, the sacrificial backside GaAs ($\sim$ 520-$\mu$m thick in our case) and the etch-stop layer are removed, and the final backside metallization is performed (Fig. \ref{fig:fig1}d). First, most ($\sim$470 $\mu$m) of the sacrificial GaAs is removed with mechanical lapping, and the remaining GaAs ($\sim$50 $\mu$m) and the etch-stop layer are selectively wet-etched using a citric acid solution and a diluted HF solution, respectively. This method of backside GaAs removal is well-established, and further details are provided in Ref.\onlinecite{Blount1996} and in our supplementary material.

After the removal of the sacrificial GaAs and the etch stop layer, the back side of the GaAs/AlGaAs epitaxial layer is now exposed with a mirror finish and ready for the final backside metallization. Because the cured epoxy is chemically resistant to the selective wet etches, the epoxy on the excess area of the substrate (edge beads) sits higher than the thinned-down GaAs/AlGaAs epitaxial layers, and must be removed for accurate patterning results. At this stage, the thin (< 1 $\mu$m) GaAs/AlGaAs epitaxial layers are optically semi-transparent, allowing accurate alignment for the backside patterns using the alignment marks defined on the front (now buried) side of the QW die. Because both the buried frontside contacts and the exposed backside contacts are routed to the bonding pads defined on the excess area of the substrate, wire bonds can be directly made on these bonding pads for easy and reliable electrical access to the contacts (Fig. \ref{fig:fig1}d).

Using the IBASE process, we fabricated ungated, single-gated, and dual-gated Hall bars, as well as small ($\sim$ 5 $\mu$m $\times$ 5 $\mu$m) dual-gate structures for TACIT detectors, and now present their characterization results. For all devices, the backside GaAs and the etch-stop layers were completely removed, leaving only 660-nm thick QW structure (as illustrated in Fig. \ref{fig:fig1}d) on the 2DEG mesas of the devices. The QW structure hosted a modulation-doped, single, 40-nm square QW. We fabricated the ungated Hall bar from QW1 with n\textsubscript{s} $=2.5 \times 10^{11}$ cm\textsuperscript{-2} (with $\mu = 7.3\times 10^{6}$ cm\textsuperscript{2}V\textsuperscript{-1}s\textsuperscript{-1} at 2 K) and the rest of the devices from QW2 that had a slightly smaller charge density of n\textsubscript{s} $=2.1 \times 10^{11}$ cm\textsuperscript{-2} (with $\mu = 7.0\times 10^{6}$ cm\textsuperscript{2}V\textsuperscript{-1}s\textsuperscript{-1} at 2 K).

\begin{figure}
\includegraphics[width=0.5\textwidth]{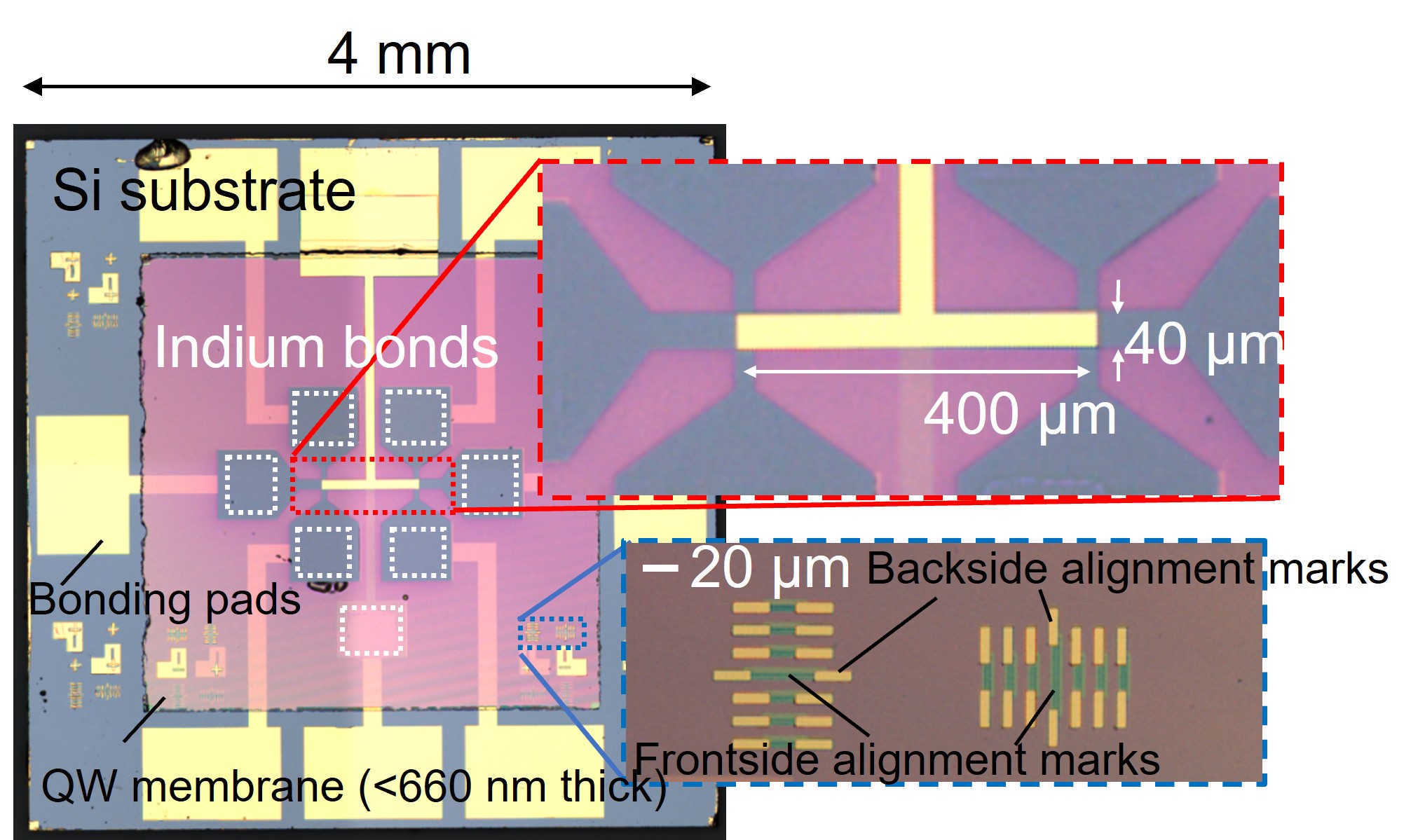}
\caption{An optical microscope image showing a dual-gated Hall bar fabricated with the IBASE process. The top inset shows the 2DEG channel (40 $\mu$m wide and 400 $\mu$m long) sandwiched by the dual-gate structure. The bottom inset shows the alignment marks used for the top gate pattern. } 

\label{fig:fig2}
\end{figure}

First of all, we present the excellent surface quality and good alignment accuracy possible with the IBASE process. Figure \ref{fig:fig2} shows the dual-gated Hall bar device fabricated with the IBASE process. The image shows the thinned-down (< 660-nm thick on the mesa-etched area) QW membrane (pink in color)  supported by the Si host substrate. The 2DEG mesa for the Hall bar (grey in color) was defined on the front (now buried) side of the QW membrane but visible on the back (now exposed) side at the end of the IBASE process. The top inset shows the 2DEG channel with the dual-gate structure enclosing the 660-nm thick 2DEG channel. As shown in the bottom inset, the optical semi-transparency of the thin QW membrane makes the alignment marks defined on the front side visible from the back side, allowing easy and accurate alignment for the backside pattern. The frontside contacts (six Ohmic contacts and the bottom Schottky gate) were flip-chip bonded to the matching electrodes on the host substrate with indium bonds (marked by white dotted rectangles). These electrodes are routed to the bonding pads defined on the excess area of the host substrate for wire-bonding. 

\begin{figure}
\includegraphics[width=0.45\textwidth]{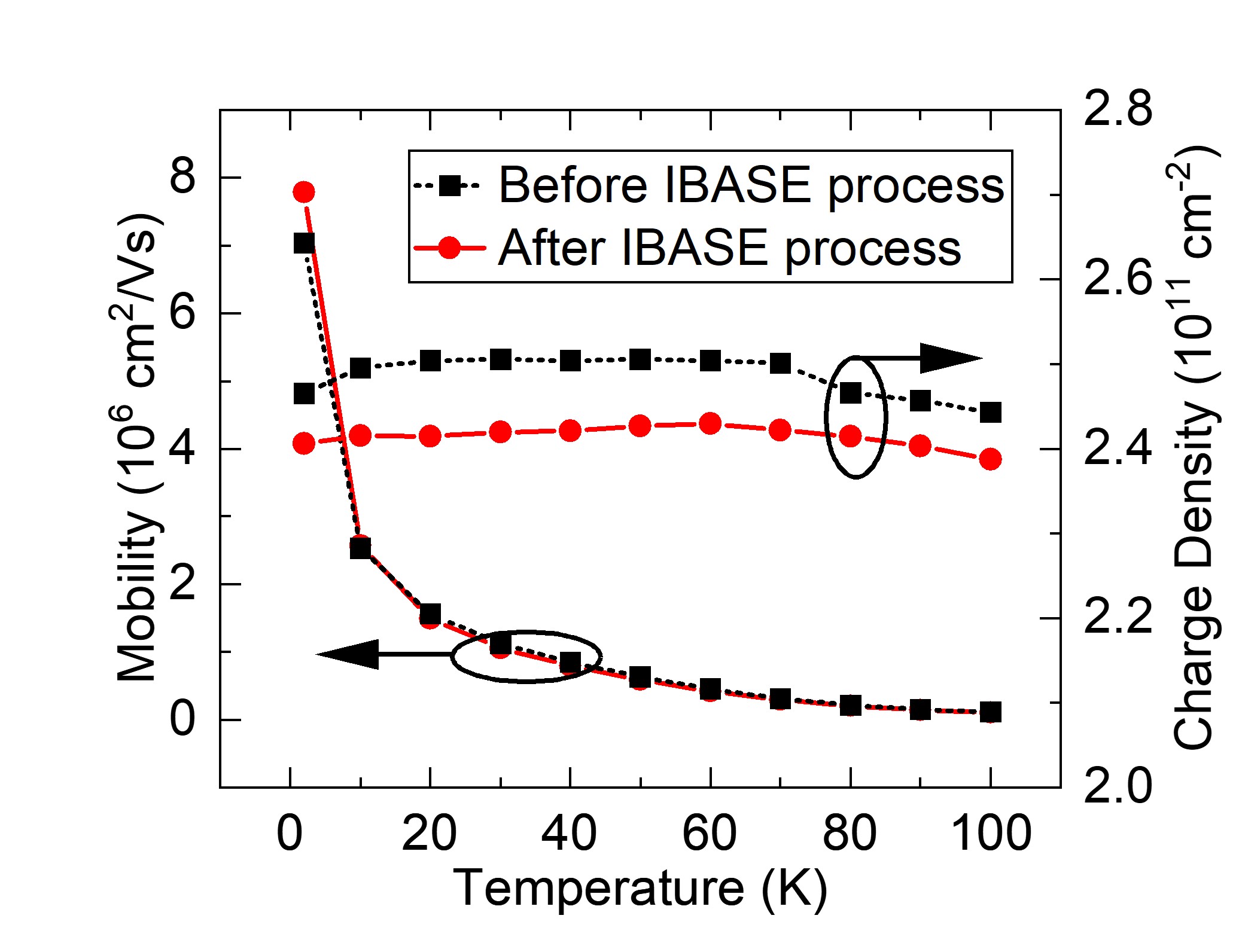}
\caption{\label{fig:fig3} Mobility and charge density characterization in an ungated Hall bar before and after the IBASE process.}
\end{figure}

Next, to verify the compatibility of our gating technique with high-mobility 2DEG, we measured the mobility and charge density in the \textit{ungated} Hall bar before and after the IBASE process (Fig. \ref{fig:fig3}). The ungated Hall bar had the same dimensions as the dual-gated Hall bar shown in Fig.\ref{fig:fig2} but lacked the gate structure on both sides of the QW structure. For the characterization, we performed standard Hall measurements (in the dark) on the same device before and after the IBASE process. As shown in Fig.\ref{fig:fig3}, both the high mobility and most (>95 \%) of the charge density were preserved after the IBASE process, confirming the good compatibility of our gating technique with the high-mobility 2DEG systems in GaAs/AlGaAs epitaxial layers. We point out that, near 2 K, we observed an increase in the mobility after the IBASE process likely due to a small fluctuation in the temperature and a slight change in the built-in electric field in the QW. 

\begin{figure}
\includegraphics[width=0.45\textwidth]{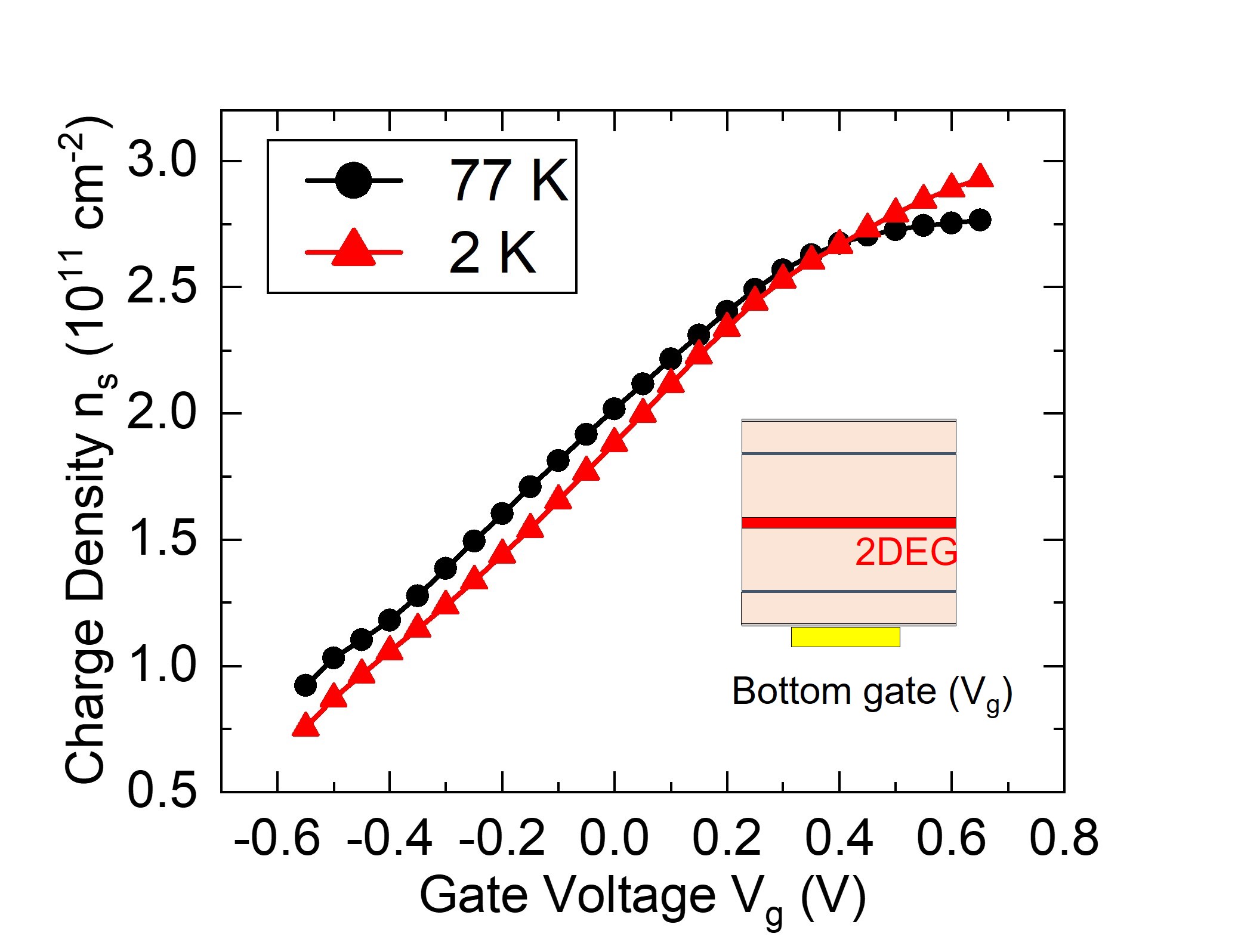}
\caption{\label{fig:fig4} Charge density as a function of gate bias in a single-gated Hall bar fabricated with the IBASE process. The schematic for the gate structure is shown in the inset. The depletion voltage is estimated to be $\sim -1$ V by extrapolating the data.}
\end{figure}

In addition, to characterize the charge tunability of an individual gate structure, we fabricated the \textit{single-gated} Hall bar and measured its charge tunability (Fig. \ref{fig:fig4}). The single-gated Hall bar had the same dimension as the dual-gated Hall bar shown in Fig. \ref{fig:fig2}, but had only one Schottky gate as the bottom gate on the front (buried) side (see the inset) of the QW membrane. To characterize the charge tunability, we swept the electrostatic gate bias from -0.6 V to 0.6 V with a step of 0.05 V, and measured the charge density using the Hall measurements (in the dark) at each voltage bias. Figure \ref{fig:fig4} shows the linear charge control achieved with small (< 1 V) DC gate biases at 77 K and 2 K. At both temperatures, the charge density was completely depleted at V\textsubscript{g} $ \sim -1 $V and started to saturate at V\textsubscript{g} $ \sim 0.4 $V. The slope in the linear region is $2.0(1) \times$ cm\textsuperscript{-2}/V for both temperatures, which agrees well with the predicted value based on a simple parallel capacitor model, in which the charge density $n_{s}$ is given by 
\begin{equation}
\label{ns}
    n_{s} = n_{0}+\frac{c}{e}V_{g}
\end{equation}
where $n_{0}$ is the intrinsic charge density, $c$ the unit-area capacitance formed by a gate metal and a 2DEG, $e$ the electron charge, and $c/e = \epsilon_{r}\epsilon_{0}/(ed) \sim 2.15 \times 10^{11}$ cm\textsuperscript{-2}/V is the slope with  the relative permittivity $\epsilon_{r}$ for GaAs ($\epsilon_{r}$ = 12.9), vacuum permittivity $\epsilon_{0}$, and the separation distance between the gate and the 2DEG $d$ for our QW ($ d = 330$ nm).

\begin{figure}
\includegraphics[width=0.5\textwidth]{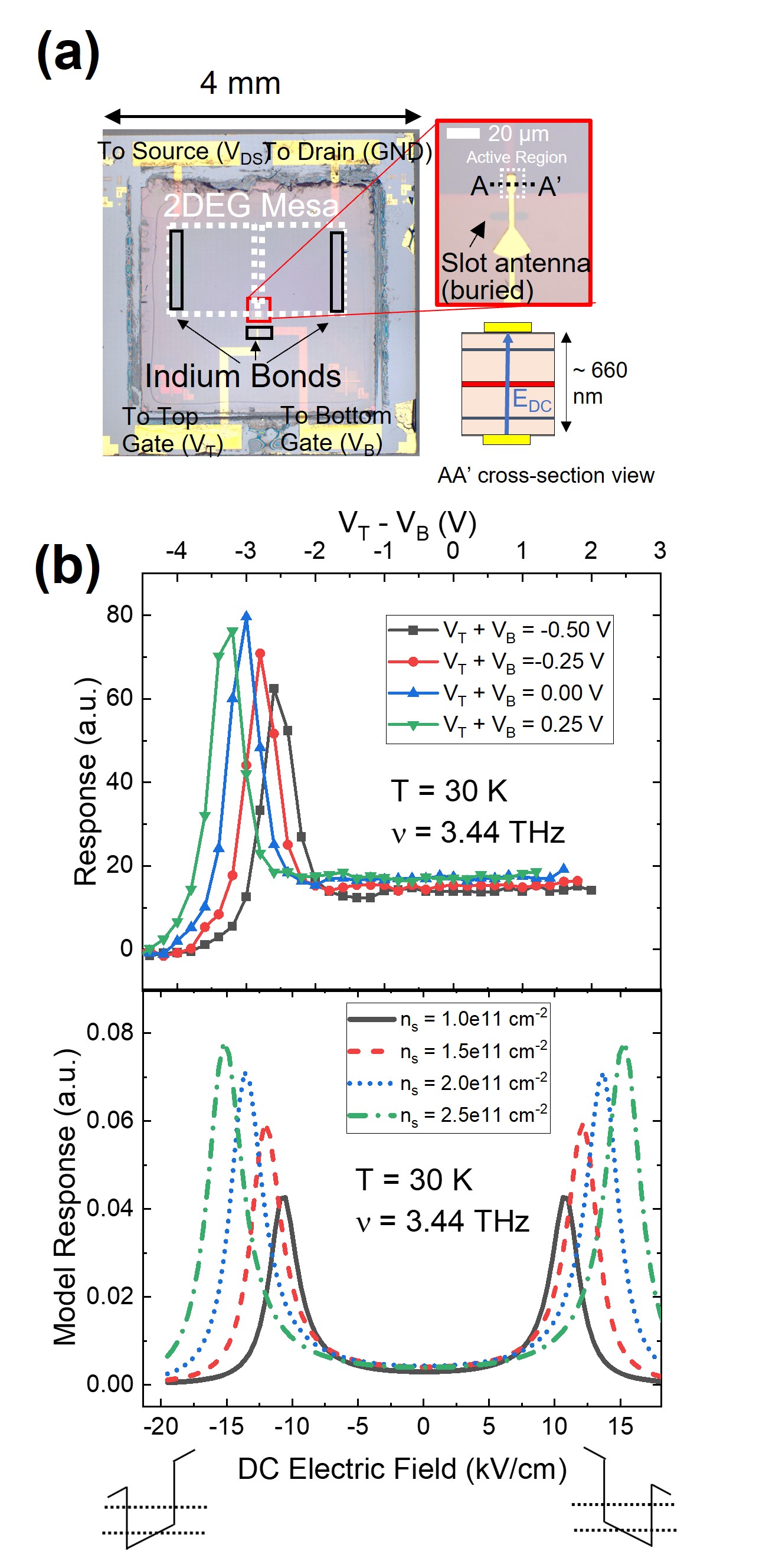}
\caption{\label{fig:fig5} Dual-gate structure in a prototype TACIT detector and its dual-gate operation (a) An optical microscope image showing the fabricated TACIT device with the dual-gate structure (b) Experimental and model responses of the device at 3.44 THz at 30 K. The schematics on the bottom illustrate the tilting of the well potential with the DC electric field.}
\end{figure}

Finally, as a practical device application of our gating technique, we fabricated a dual-gate structure for a prototype TACIT detector\cite{Yoo2020} and investigated its dual-gating operation (Fig. \ref{fig:fig5}). Figure \ref{fig:fig5}a shows the fabricated device with the dual-gate structure. The dual-gate structure enclosed a small ($\sim$ 5 $\mu$m $\times$ 5 $\mu$m) active region of the device where THz absorption occurs via intersubband transitions. Besides the dual-gate structure, two 1-mm wide Ohmic contacts (source and drain) were defined for the read-out of the device response.\footnote{The wide contacts were to minimize the contact resistances.} As shown in the top right inset, a slot antenna along with a THz stub and choke was integrated with the dual-gate metallization for efficient THz coupling. For our 40-nm QW, the bare intersubband absorption frequency (assuming negligible charge density and flat band conditions) lies near 2.5 THz, but could be tuned in a wide range (2.5-5 THz) by varying the charge density and the DC electric field oriented along the growth direction of the QW (see the bottom right inset).\cite{Williams2001, Yoo2020} Further details on the device fabrication and operation are provided in Ref. \onlinecite{Yoo2020} and in the supplementary material. 

To investigate the dual-gating operation, we first fixed the charge density and varied the DC electric field while we monitored the direct detection response of the device. For the independent control of the two tuning parameters, we defined the sum voltage $V_{S} = V_{T} + V_{B}$ and the difference voltage $V_{D} = V_{T}-V_{B}$ where $V_{T(B)}$ is the top (bottom) gate voltage relative to the grounded drain of the device (Fig. \ref{fig:fig5}a). Using these voltage values, the total charge density $n_{s}$ is tuned with $V_{S}$ according to  Eq. \ref{ns} with $V_{S}$ replacing $V_{g}$, and the DC electric field $E_{DC}$ is tuned by $V_{D}$. This allowed independent tuning of $n_{s}$ and $E_{DC}$ (e.g. fixing $n_{s}$ and sweeping $E_{DC}$ by varying $V_{T}$ and $V_{B}$ simultaneously to fix $V_{S}$ constant and sweep $V_{D}$), which is critical for the precise and accurate tuning of the intersubband absorption frequency.

For the device response, we measured the change in the in-plane resistance of the 2DEG (across the source and the drain) by applying a small (< 10$\mu$A) constant DC current bias and by measuring the change in the source-drain voltage ($V_{DS}$) in response to a monochromatic radiation at 3.44 THz at 30 K. The THz radiation, provided by a quantum-cascade vertical-external-cavity surface-emitting laser (QC-VECSEL) operated at 77 K,\cite{Xu2017, Curwen2019} was quasi-optically coupled to the device, and a lock-in amplifier was used to measure the voltage response.\footnote{Further detail on the experimental setup is provided in the supplementary material.} 

The direct detection response of the TACIT device along with the model response as a function of DC electric field at four different charge densities is shown in Fig.\ref{fig:fig5}b. The legends show the values of the sum voltages ($V_{S}$) and the corresponding charge densities converted using Eq.\ref{ns}. For the model response curves, we calculated the impedance matching efficiency between the antenna and the active region of the device based on the THz impedance model for the intersubband absorption.\cite{Yoo2022} To compare the data with the model, we converted the difference voltage values ($V_{D}$) in the data (indicated in the top x-axis) to the DC electric field values (indicated in the bottom x-axis) based on the calibration done at 3.11 THz.\footnote{For further details on the modeling and the DC electric field calibration, see the supplementary material.} As shown in Fig. \ref{fig:fig5}b, we observed that the device response peaks near the $V_{D}$ values where the intersubband absorption frequency of our 40-nm QW matches the incoming THz radiation at 3.44 THz. As predicted in the model response, the peak response became higher, and the location of peak absorption shifted outward with an increasing charge density. These observations confirm the independent tuning of the charge density and the DC electric field with the fabricated dual-gate structure. We note that, in the experimental data, we could not observe the second peaks in the device response predicted at $V_{D} > 1 V$ in the model due to the gate breakdown.

In summary, we demonstrated an alternative, reliable flip-chip technique that allows dual-side processing of thin (< 1 $\mu$m) high-mobility GaAs/AlGaAs epitaxial layers without complicated backside processing for reliable electrical contacts. The technique preserves both the high mobility and most (>95 \%) of the charge density of the 2DEG systems in the GaAs/AlGaAs epitaxial layers, and allows linear control of the charge density with small (< 1V) gate bias. Lastly, we demonstrated the independent control of the charge density and the DC electric field set between the dual gates by tuning the THz intersubband absorption of our prototype TACIT device with the dual-gate structure fabricated with the IBASE process. We believe that our gating technique is important and easily applicable for both practical device applications and fundamental study of quantum transports using high-mobility GaAs/AlGaAs heterostructures. 


\begin{acknowledgments}
This research was carried out in part at the Jet Propulsion Laboratory, California Institute of Technology, under a contract with the National Aeronautics and Space Administration. The device fabrication for this work was performed in the UCSB Nanofabrication Facility, an open access laboratory. We thank Dr. Mengchen Huang for his contribution to the early development of the IBASE processs, and Dr. Brian Thibeault and Dr. Demis D. John for their help in discussing the flip-chip processing steps. We thank Prof. Benjamin Williams for his help in providing the QC-VECSEL at 3.44 THz. The C++ codes used for the model responses of TACIT detectors were written by Dr. Bryan Galdrikian and modified by Dr. Chris Morris. The work at UCSB was supported by NASA PICASSO program via a contract with JPL. The UCSB MRL Shared Experimental Facilities are supported by the MRSEC Program of the NSF under Award No. DMR 1720256; a member of the NSF-funded Materials Research Facilities Network (www.mrfn.org). 

\end{acknowledgments}

\bibliography{IBASE}

\begin{thebibliography}{21}%
\makeatletter
\providecommand \@ifxundefined [1]{%
 \@ifx{#1\undefined}
}%
\providecommand \@ifnum [1]{%
 \ifnum #1\expandafter \@firstoftwo
 \else \expandafter \@secondoftwo
 \fi
}%
\providecommand \@ifx [1]{%
 \ifx #1\expandafter \@firstoftwo
 \else \expandafter \@secondoftwo
 \fi
}%
\providecommand \natexlab [1]{#1}%
\providecommand \enquote  [1]{``#1''}%
\providecommand \bibnamefont  [1]{#1}%
\providecommand \bibfnamefont [1]{#1}%
\providecommand \citenamefont [1]{#1}%
\providecommand \href@noop [0]{\@secondoftwo}%
\providecommand \href [0]{\begingroup \@sanitize@url \@href}%
\providecommand \@href[1]{\@@startlink{#1}\@@href}%
\providecommand \@@href[1]{\endgroup#1\@@endlink}%
\providecommand \@sanitize@url [0]{\catcode `\\12\catcode `\$12\catcode
  `\&12\catcode `\#12\catcode `\^12\catcode `\_12\catcode `\%12\relax}%
\providecommand \@@startlink[1]{}%
\providecommand \@@endlink[0]{}%
\providecommand \url  [0]{\begingroup\@sanitize@url \@url }%
\providecommand \@url [1]{\endgroup\@href {#1}{\urlprefix }}%
\providecommand \urlprefix  [0]{URL }%
\providecommand \Eprint [0]{\href }%
\providecommand \doibase [0]{http://dx.doi.org/}%
\providecommand \selectlanguage [0]{\@gobble}%
\providecommand \bibinfo  [0]{\@secondoftwo}%
\providecommand \bibfield  [0]{\@secondoftwo}%
\providecommand \translation [1]{[#1]}%
\providecommand \BibitemOpen [0]{}%
\providecommand \bibitemStop [0]{}%
\providecommand \bibitemNoStop [0]{.\EOS\space}%
\providecommand \EOS [0]{\spacefactor3000\relax}%
\providecommand \BibitemShut  [1]{\csname bibitem#1\endcsname}%
\let\auto@bib@innerbib\@empty
\bibitem [{\citenamefont {Eisenstein}, \citenamefont {Pfeiffer},\ and\
  \citenamefont {West}(1990)}]{Eisenstein1990}%
  \BibitemOpen
  \bibfield  {author} {\bibinfo {author} {\bibfnamefont {J.~P.}\ \bibnamefont
  {Eisenstein}}, \bibinfo {author} {\bibfnamefont {L.~N.}\ \bibnamefont
  {Pfeiffer}}, \ and\ \bibinfo {author} {\bibfnamefont {K.~W.}\ \bibnamefont
  {West}},\ }\href {\doibase 10.1063/1.103882} {\bibfield  {journal} {\bibinfo
  {journal} {Applied Physics Letters}\ }\textbf {\bibinfo {volume} {57}},\
  \bibinfo {pages} {2324} (\bibinfo {year} {1990})}\BibitemShut {NoStop}%
\bibitem [{\citenamefont {Croxall}\ \emph {et~al.}(2013)\citenamefont
  {Croxall}, \citenamefont {Zheng}, \citenamefont {Sfigakis}, \citenamefont
  {{Das Gupta}}, \citenamefont {Farrer}, \citenamefont {Nicoll}, \citenamefont
  {Beere},\ and\ \citenamefont {Ritchie}}]{Croxall2013}%
  \BibitemOpen
  \bibfield  {author} {\bibinfo {author} {\bibfnamefont {A.~F.}\ \bibnamefont
  {Croxall}}, \bibinfo {author} {\bibfnamefont {B.}~\bibnamefont {Zheng}},
  \bibinfo {author} {\bibfnamefont {F.}~\bibnamefont {Sfigakis}}, \bibinfo
  {author} {\bibfnamefont {K.}~\bibnamefont {{Das Gupta}}}, \bibinfo {author}
  {\bibfnamefont {I.}~\bibnamefont {Farrer}}, \bibinfo {author} {\bibfnamefont
  {C.~A.}\ \bibnamefont {Nicoll}}, \bibinfo {author} {\bibfnamefont {H.~E.}\
  \bibnamefont {Beere}}, \ and\ \bibinfo {author} {\bibfnamefont {D.~A.}\
  \bibnamefont {Ritchie}},\ }\href@noop {} {\bibfield  {journal} {\bibinfo
  {journal} {Applied Physics Letters}\ }\textbf {\bibinfo {volume} {102}}
  (\bibinfo {year} {2013})}\BibitemShut {NoStop}%
\bibitem [{\citenamefont {Sherwin}(1999)}]{Sherwin1997}%
  \BibitemOpen
  \bibfield  {author} {\bibinfo {author} {\bibfnamefont {M.~S.}\ \bibnamefont
  {Sherwin}},\ }\href {https://patents.google.com/patent/US5914497} {}\bibinfo
  {howpublished} {U.S. Patent No.5,914,497} (\bibinfo {year} {22 Jun.
  1999})\BibitemShut {NoStop}%
\bibitem [{\citenamefont {Sherwin}\ \emph {et~al.}(2002)\citenamefont
  {Sherwin}, \citenamefont {Cates}, \citenamefont {Serapiglia}, \citenamefont
  {Dora}, \citenamefont {Williams}, \citenamefont {Maranowski}, \citenamefont
  {Gossard},\ and\ \citenamefont {McGrath}}]{Sherwin2002}%
  \BibitemOpen
  \bibfield  {author} {\bibinfo {author} {\bibfnamefont {M.~S.}\ \bibnamefont
  {Sherwin}}, \bibinfo {author} {\bibfnamefont {C.}~\bibnamefont {Cates}},
  \bibinfo {author} {\bibfnamefont {B.}~\bibnamefont {Serapiglia}}, \bibinfo
  {author} {\bibfnamefont {Y.}~\bibnamefont {Dora}}, \bibinfo {author}
  {\bibfnamefont {J.~B.}\ \bibnamefont {Williams}}, \bibinfo {author}
  {\bibfnamefont {K.~D.}\ \bibnamefont {Maranowski}}, \bibinfo {author}
  {\bibfnamefont {A.~C.}\ \bibnamefont {Gossard}}, \ and\ \bibinfo {author}
  {\bibfnamefont {W.~R.}\ \bibnamefont {McGrath}},\ }\href@noop {} {}\bibinfo
  {howpublished} {Proceedings of Far-IR, Submm, and mm Detector Technology
  Workshop, Monterey, CA} (\bibinfo {year} {2002}),\ \bibinfo {note} {e-print
  available at https://arxiv.org/abs/1909.10664}\BibitemShut {NoStop}%
\bibitem [{\citenamefont {Yoo}\ \emph {et~al.}(2020)\citenamefont {Yoo},
  \citenamefont {Huang}, \citenamefont {Kawamura}, \citenamefont {West},
  \citenamefont {Pfeiffer}, \citenamefont {Karasik},\ and\ \citenamefont
  {Sherwin}}]{Yoo2020}%
  \BibitemOpen
  \bibfield  {author} {\bibinfo {author} {\bibfnamefont {C.}~\bibnamefont
  {Yoo}}, \bibinfo {author} {\bibfnamefont {M.}~\bibnamefont {Huang}}, \bibinfo
  {author} {\bibfnamefont {J.~H.}\ \bibnamefont {Kawamura}}, \bibinfo {author}
  {\bibfnamefont {K.~W.}\ \bibnamefont {West}}, \bibinfo {author}
  {\bibfnamefont {L.~N.}\ \bibnamefont {Pfeiffer}}, \bibinfo {author}
  {\bibfnamefont {B.~S.}\ \bibnamefont {Karasik}}, \ and\ \bibinfo {author}
  {\bibfnamefont {M.~S.}\ \bibnamefont {Sherwin}},\ }\href@noop {} {\bibfield
  {journal} {\bibinfo  {journal} {Applied Physics Letters}\ }\textbf {\bibinfo
  {volume} {116}} (\bibinfo {year} {2020})}\BibitemShut {NoStop}%
\bibitem [{\citenamefont {Helm}(1999)}]{Helm1999}%
  \BibitemOpen
  \bibfield  {author} {\bibinfo {author} {\bibfnamefont {M.}~\bibnamefont
  {Helm}},\ }\href {\doibase 10.1016/S0080-8784(08)60304-X} {\bibfield
  {journal} {\bibinfo  {journal} {Semiconductors and Semimetals}\ }\textbf
  {\bibinfo {volume} {62}},\ \bibinfo {pages} {1} (\bibinfo {year}
  {1999})}\BibitemShut {NoStop}%
\bibitem [{\citenamefont {Linfield}\ \emph {et~al.}(1993)\citenamefont
  {Linfield}, \citenamefont {Jones}, \citenamefont {Ritchie}, \citenamefont
  {Hamilton},\ and\ \citenamefont {Iredale}}]{Linfield1993}%
  \BibitemOpen
  \bibfield  {author} {\bibinfo {author} {\bibfnamefont {E.~H.}\ \bibnamefont
  {Linfield}}, \bibinfo {author} {\bibfnamefont {G.~A.}\ \bibnamefont {Jones}},
  \bibinfo {author} {\bibfnamefont {D.~A.}\ \bibnamefont {Ritchie}}, \bibinfo
  {author} {\bibfnamefont {A.~R.}\ \bibnamefont {Hamilton}}, \ and\ \bibinfo
  {author} {\bibfnamefont {N.}~\bibnamefont {Iredale}},\ }\href {\doibase
  10.1016/0022-0248(93)90573-F} {\bibfield  {journal} {\bibinfo  {journal}
  {Journal of Crystal Growth}\ }\textbf {\bibinfo {volume} {127}},\ \bibinfo
  {pages} {41} (\bibinfo {year} {1993})}\BibitemShut {NoStop}%
\bibitem [{\citenamefont {Evans}\ \emph {et~al.}(1994)\citenamefont {Evans},
  \citenamefont {Grimshaw}, \citenamefont {Burroughes}, \citenamefont
  {Leadbeater}, \citenamefont {Tribble}, \citenamefont {Ritchie}, \citenamefont
  {Jones},\ and\ \citenamefont {Pepper}}]{Evans1994}%
  \BibitemOpen
  \bibfield  {author} {\bibinfo {author} {\bibfnamefont {R.~J.}\ \bibnamefont
  {Evans}}, \bibinfo {author} {\bibfnamefont {M.~P.}\ \bibnamefont {Grimshaw}},
  \bibinfo {author} {\bibfnamefont {J.~H.}\ \bibnamefont {Burroughes}},
  \bibinfo {author} {\bibfnamefont {M.~L.}\ \bibnamefont {Leadbeater}},
  \bibinfo {author} {\bibfnamefont {M.~J.}\ \bibnamefont {Tribble}}, \bibinfo
  {author} {\bibfnamefont {D.~A.}\ \bibnamefont {Ritchie}}, \bibinfo {author}
  {\bibfnamefont {G.~A.}\ \bibnamefont {Jones}}, \ and\ \bibinfo {author}
  {\bibfnamefont {M.}~\bibnamefont {Pepper}},\ }\href {\doibase
  10.1063/1.112824} {\bibfield  {journal} {\bibinfo  {journal} {Applied Physics
  Letters}\ }\textbf {\bibinfo {volume} {65}},\ \bibinfo {pages} {1943}
  (\bibinfo {year} {1994})}\BibitemShut {NoStop}%
\bibitem [{\citenamefont {Brown}\ \emph {et~al.}(1994)\citenamefont {Brown},
  \citenamefont {Linfield}, \citenamefont {Ritchie}, \citenamefont {Jones},
  \citenamefont {Grimshaw},\ and\ \citenamefont {Pepper}}]{Brown1994}%
  \BibitemOpen
  \bibfield  {author} {\bibinfo {author} {\bibfnamefont {K.~M.}\ \bibnamefont
  {Brown}}, \bibinfo {author} {\bibfnamefont {E.~H.}\ \bibnamefont {Linfield}},
  \bibinfo {author} {\bibfnamefont {D.~A.}\ \bibnamefont {Ritchie}}, \bibinfo
  {author} {\bibfnamefont {G.~A.}\ \bibnamefont {Jones}}, \bibinfo {author}
  {\bibfnamefont {M.~P.}\ \bibnamefont {Grimshaw}}, \ and\ \bibinfo {author}
  {\bibfnamefont {M.}~\bibnamefont {Pepper}},\ }\href {\doibase
  10.1063/1.111768} {\bibfield  {journal} {\bibinfo  {journal} {Applied Physics
  Letters}\ }\textbf {\bibinfo {volume} {64}},\ \bibinfo {pages} {1827}
  (\bibinfo {year} {1994})}\BibitemShut {NoStop}%
\bibitem [{\citenamefont {Berl}\ \emph {et~al.}(2016)\citenamefont {Berl},
  \citenamefont {Tiemann}, \citenamefont {Dietsche}, \citenamefont {Karl},\
  and\ \citenamefont {Wegscheider}}]{Berl2016}%
  \BibitemOpen
  \bibfield  {author} {\bibinfo {author} {\bibfnamefont {M.}~\bibnamefont
  {Berl}}, \bibinfo {author} {\bibfnamefont {L.}~\bibnamefont {Tiemann}},
  \bibinfo {author} {\bibfnamefont {W.}~\bibnamefont {Dietsche}}, \bibinfo
  {author} {\bibfnamefont {H.}~\bibnamefont {Karl}}, \ and\ \bibinfo {author}
  {\bibfnamefont {W.}~\bibnamefont {Wegscheider}},\ }\href
  {http://dx.doi.org/10.1063/1.4945090} {\bibfield  {journal} {\bibinfo
  {journal} {Applied Physics Letters}\ }\textbf {\bibinfo {volume} {108}}
  (\bibinfo {year} {2016})}\BibitemShut {NoStop}%
\bibitem [{\citenamefont {Weckwerth}\ \emph {et~al.}(1996)\citenamefont
  {Weckwerth}, \citenamefont {Simmons}, \citenamefont {Harff}, \citenamefont
  {Sherwin}, \citenamefont {Blount}, \citenamefont {Baca},\ and\ \citenamefont
  {Chui}}]{Blount1996}%
  \BibitemOpen
  \bibfield  {author} {\bibinfo {author} {\bibfnamefont {M.~V.}\ \bibnamefont
  {Weckwerth}}, \bibinfo {author} {\bibfnamefont {J.~A.}\ \bibnamefont
  {Simmons}}, \bibinfo {author} {\bibfnamefont {N.~E.}\ \bibnamefont {Harff}},
  \bibinfo {author} {\bibfnamefont {M.~E.}\ \bibnamefont {Sherwin}}, \bibinfo
  {author} {\bibfnamefont {M.~A.}\ \bibnamefont {Blount}}, \bibinfo {author}
  {\bibfnamefont {W.~E.}\ \bibnamefont {Baca}}, \ and\ \bibinfo {author}
  {\bibfnamefont {H.~C.}\ \bibnamefont {Chui}},\ }\href {\doibase
  10.1006/spmi.1996.0115} {\bibfield  {journal} {\bibinfo  {journal}
  {Superlattices and Microstructures}\ }\textbf {\bibinfo {volume} {20}},\
  \bibinfo {pages} {561} (\bibinfo {year} {1996})}\BibitemShut {NoStop}%
\bibitem [{\citenamefont {Croxall}\ \emph {et~al.}(2008)\citenamefont
  {Croxall}, \citenamefont {{Das Gupta}}, \citenamefont {Nicoll}, \citenamefont
  {Thangaraj}, \citenamefont {Farrer}, \citenamefont {Ritchie},\ and\
  \citenamefont {Pepper}}]{Croxall2008}%
  \BibitemOpen
  \bibfield  {author} {\bibinfo {author} {\bibfnamefont {A.~F.}\ \bibnamefont
  {Croxall}}, \bibinfo {author} {\bibfnamefont {K.}~\bibnamefont {{Das
  Gupta}}}, \bibinfo {author} {\bibfnamefont {C.~A.}\ \bibnamefont {Nicoll}},
  \bibinfo {author} {\bibfnamefont {M.}~\bibnamefont {Thangaraj}}, \bibinfo
  {author} {\bibfnamefont {I.}~\bibnamefont {Farrer}}, \bibinfo {author}
  {\bibfnamefont {D.~A.}\ \bibnamefont {Ritchie}}, \ and\ \bibinfo {author}
  {\bibfnamefont {M.}~\bibnamefont {Pepper}},\ }\href@noop {} {\bibfield
  {journal} {\bibinfo  {journal} {Journal of Applied Physics}\ }\textbf
  {\bibinfo {volume} {104}},\ \bibinfo {pages} {1} (\bibinfo {year}
  {2008})}\BibitemShut {NoStop}%
\bibitem [{\citenamefont {Gupta}\ \emph {et~al.}(2012)\citenamefont {Gupta},
  \citenamefont {Croxall}, \citenamefont {Mak}, \citenamefont {Beere},
  \citenamefont {Nicoll}, \citenamefont {Farrer}, \citenamefont {Sfigakis},\
  and\ \citenamefont {Ritchie}}]{Gupta2012}%
  \BibitemOpen
  \bibfield  {author} {\bibinfo {author} {\bibfnamefont {K.~D.}\ \bibnamefont
  {Gupta}}, \bibinfo {author} {\bibfnamefont {A.~F.}\ \bibnamefont {Croxall}},
  \bibinfo {author} {\bibfnamefont {W.~Y.}\ \bibnamefont {Mak}}, \bibinfo
  {author} {\bibfnamefont {H.~E.}\ \bibnamefont {Beere}}, \bibinfo {author}
  {\bibfnamefont {C.~A.}\ \bibnamefont {Nicoll}}, \bibinfo {author}
  {\bibfnamefont {I.}~\bibnamefont {Farrer}}, \bibinfo {author} {\bibfnamefont
  {F.}~\bibnamefont {Sfigakis}}, \ and\ \bibinfo {author} {\bibfnamefont
  {D.~A.}\ \bibnamefont {Ritchie}},\ }\href@noop {} {\bibfield  {journal}
  {\bibinfo  {journal} {Semiconductor Science and Technology}\ }\textbf
  {\bibinfo {volume} {27}} (\bibinfo {year} {2012})}\BibitemShut {NoStop}%
\bibitem [{Note1()}]{Note1}%
  \BibitemOpen
  \bibinfo {note} {While our gating technique can be applied to a variety of QW
  heterostructures including double-QWs and superlattices, we used a single,
  square QW for the fabrication of the devices presented in this
  Letter.}\BibitemShut {Stop}%
\bibitem [{Note2()}]{Note2}%
  \BibitemOpen
  \bibinfo {note} {The wide contacts were to minimize the contact
  resistances.}\BibitemShut {Stop}%
\bibitem [{\citenamefont {Williams}\ \emph {et~al.}(2001)\citenamefont
  {Williams}, \citenamefont {Sherwin}, \citenamefont {Maranowski},\ and\
  \citenamefont {Gossard}}]{Williams2001}%
  \BibitemOpen
  \bibfield  {author} {\bibinfo {author} {\bibfnamefont {J.~B.}\ \bibnamefont
  {Williams}}, \bibinfo {author} {\bibfnamefont {M.~S.}\ \bibnamefont
  {Sherwin}}, \bibinfo {author} {\bibfnamefont {K.~D.}\ \bibnamefont
  {Maranowski}}, \ and\ \bibinfo {author} {\bibfnamefont {A.~C.}\ \bibnamefont
  {Gossard}},\ }\href {\doibase 10.1103/PhysRevLett.87.037401} {\bibfield
  {journal} {\bibinfo  {journal} {Physical Review Letters}\ }\textbf {\bibinfo
  {volume} {87}},\ \bibinfo {pages} {37401} (\bibinfo {year}
  {2001})}\BibitemShut {NoStop}%
\bibitem [{\citenamefont {Xu}\ \emph {et~al.}(2017)\citenamefont {Xu},
  \citenamefont {Curwen}, \citenamefont {Chen}, \citenamefont {Reno},
  \citenamefont {Itoh},\ and\ \citenamefont {Williams}}]{Xu2017}%
  \BibitemOpen
  \bibfield  {author} {\bibinfo {author} {\bibfnamefont {L.}~\bibnamefont
  {Xu}}, \bibinfo {author} {\bibfnamefont {C.~A.}\ \bibnamefont {Curwen}},
  \bibinfo {author} {\bibfnamefont {D.}~\bibnamefont {Chen}}, \bibinfo {author}
  {\bibfnamefont {J.~L.}\ \bibnamefont {Reno}}, \bibinfo {author}
  {\bibfnamefont {T.}~\bibnamefont {Itoh}}, \ and\ \bibinfo {author}
  {\bibfnamefont {B.~S.}\ \bibnamefont {Williams}},\ }\href {\doibase
  10.1109/JSTQE.2017.2693024} {\bibfield  {journal} {\bibinfo  {journal} {IEEE
  Journal of Selected Topics in Quantum Electronics}\ }\textbf {\bibinfo
  {volume} {23}},\ \bibinfo {pages} {1} (\bibinfo {year} {2017})}\BibitemShut
  {NoStop}%
\bibitem [{\citenamefont {Curwen}, \citenamefont {Reno},\ and\ \citenamefont
  {Williams}(2019)}]{Curwen2019}%
  \BibitemOpen
  \bibfield  {author} {\bibinfo {author} {\bibfnamefont {C.~A.}\ \bibnamefont
  {Curwen}}, \bibinfo {author} {\bibfnamefont {J.~L.}\ \bibnamefont {Reno}}, \
  and\ \bibinfo {author} {\bibfnamefont {B.~S.}\ \bibnamefont {Williams}},\
  }\href {\doibase 10.1038/s41566-019-0518-z} {\bibfield  {journal} {\bibinfo
  {journal} {Nature Photonics}\ }\textbf {\bibinfo {volume} {13}},\ \bibinfo
  {pages} {855} (\bibinfo {year} {2019})}\BibitemShut {NoStop}%
\bibitem [{Note3()}]{Note3}%
  \BibitemOpen
  \bibinfo {note} {Further detail on the experimental setup is provided in the
  supplementary material.}\BibitemShut {Stop}%
\bibitem [{\citenamefont {Yoo}\ \emph {et~al.}(2022)\citenamefont {Yoo},
  \citenamefont {West}, \citenamefont {Pfeiffer}, \citenamefont {Kawamura},
  \citenamefont {Karasik},\ and\ \citenamefont {Sherwin}}]{Yoo2022}%
  \BibitemOpen
  \bibfield  {author} {\bibinfo {author} {\bibfnamefont {C.}~\bibnamefont
  {Yoo}}, \bibinfo {author} {\bibfnamefont {K.~W.}\ \bibnamefont {West}},
  \bibinfo {author} {\bibfnamefont {L.~N.}\ \bibnamefont {Pfeiffer}}, \bibinfo
  {author} {\bibfnamefont {J.~H.}\ \bibnamefont {Kawamura}}, \bibinfo {author}
  {\bibfnamefont {B.~S.}\ \bibnamefont {Karasik}}, \ and\ \bibinfo {author}
  {\bibfnamefont {M.~S.}\ \bibnamefont {Sherwin}},\ }\href@noop {} {}\bibinfo
  {howpublished} {Proceedings SPIE, Millimeter, Submillimeter, and Far-Infrared
  Detectors and Instrumentation for Astronomy XI, Montreal, Quebec, CA}
  (\bibinfo {year} {2022})\BibitemShut {NoStop}%
\bibitem [{Note4()}]{Note4}%
  \BibitemOpen
  \bibinfo {note} {For further details on the modeling and the DC electric
  field calibration, see the supplementary material.}\BibitemShut {Stop}%
\end{thebibliography}%

\end{document}


\title[Supplementary Material]{Supplementary Material for Indium-Bond-And-Stop-Etch (IBASE) Technique for Dual-side Processing of Thin High-mobility GaAs/AlGaAs Epitaxial Layers}

\author{C. Yoo}
\affiliation{Physics Department and Institute for Terahertz Science and Technology, University of California, Santa Barbara, California 93106 USA}
 
\author{M. Huang}
\affiliation{Physics Department and Institute for Terahertz Science and Technology, University of California, Santa Barbara, California 93106 USA}

\author{K. W. West}
\affiliation{Department of Electrical Engineering, Princeton University, Princeton, New Jersey 08544 USA}
 
\author{L. N. Pfeiffer}
\affiliation{Department of Electrical Engineering, Princeton University, Princeton, New Jersey 08544 USA}
 
\author{C. A. Curwen}
\affiliation{Jet Propulsion Laboratory, California Institute of Technology, Pasadena, California 91109 USA}

\author{J. H. Kawamura}
\affiliation{Jet Propulsion Laboratory, California Institute of Technology, Pasadena, California 91109 USA}

\author{B. S. Karasik}
\affiliation{Jet Propulsion Laboratory, California Institute of Technology, Pasadena, California 91109 USA}
 
\author{M. S. Sherwin\textsuperscript{*}}
 \email{sherwin@ucsb.edu}
\affiliation{Physics Department and Institute for Terahertz Science and Technology, University of California, Santa Barbara, California 93106 USA}

\maketitle

\section{MBE Sample Growth}

Our QW sample was grown by molecular beam epitaxy (MBE) on a 520-$\mu$m thick intrinsic GaAs substrate (2-inch in diameter). The exact growth sequence of our QW sample is the following: a 500-nm GaAs buffer layer and a smoothing layer consisting of 100 periods of 10-nm Al\textsubscript{0.3}Ga\textsubscript{0.7}As and 3-nm GaAs, a 500-nm  Al\textsubscript{0.74}Ga\textsubscript{0.26}As layer as an etch-stop layer, a 10-nm GaAs layer for a cap layer for the QW structure, the first barrier layer consisting of 200-nm Al\textsubscript{0.3}Ga\textsubscript{0.7}As, a Si modulation doping layer, and 100-nm Al\textsubscript{0.3}Ga\textsubscript{0.7}, 40-nm GaAs QW layer, the second barrier layer consisting of a 100-nm Al\textsubscript{0.3}Ga\textsubscript{0.7}As layer, a Si modulation doping layer, a 200-nm Al\textsubscript{0.3}Ga\textsubscript{0.7}As layer, and finally another 10-nm GaAs cap layer. This created a single 40-nm wide QW with the center of the 2DEG layer 330 nm below the top surface. The QW was modulation-doped with Si doping layers placed 220-nm away from the center of the 2DEG on both sides. For the test devices demonstrated in this Letter, we used two versions of the 40-nm QW sample, labeled as QW1 and QW2, with two different Si doping levels. For QW1, the charge density $n_{s}$ and mobility $\mu$ (measured with Hall measurements in dark at 2 K) were $2.5 \times 10^{11}$ cm\textsuperscript{-2} and $7.0\times 10^{6}$ cm\textsuperscript{2}V\textsuperscript{-1}s\textsuperscript{-1}. QW2 had slightly less charge density with similar mobility. For QW2, we had $n_{s} = 2.1 \times 10^{11}$ cm\textsuperscript{-2} and $\mu = 7.3\times 10^{6}$ cm\textsuperscript{2}V\textsuperscript{-1}s\textsuperscript{-1}. 

\section{Device Fabrication}
Using the IBASE process, we fabricated ungated, single-gated, and dual-gated Hall bars, as well as small ($\sim$ 5 $\mu$m $\times$ 5 $\mu$m) dual-gate structures on 660-nm thick GaAs/AlGaAs epitaxial layers for our TACIT detectors. The ungated Hall bar device was fabricated from QW1 and the rest of the devices from QW2. For all devices, the fabrication recipes were identical, and the detailed recipe is provided below. 

\subsection{Frontside Fabrication}
For the frontside processing, we first cleaved the original QW sample wafer (2-in in diameter) into smaller pieces (12 mm x 12 mm). After cleaving, a 2DEG mesa was wet-etched to a depth of $\sim$ 400 nm using a diluted sulfuric acid solution with a mix ratio of 1:1:60 (H\textsubscript{2}SO\textsubscript{4}:H\textsubscript{2}O\textsubscript{2}:H\textsubscript{2}O). For Ohmic contacts, we performed standard UV lithography with a contact aligner and deposited Ni/AuGe/Ni/Au (4 nm/ 470 nm/ 150 nm/ 100 nm) contacts. Before the deposition, the exposed area for the contacts was cleaned with a diluted HCl solution with a mix ratio of 1:20 (HCl:H\textsubscript{2}O). After the deposition, the contacts were rapidly annealed at 370 $^\circ$C for 60 s and then at 500$^\circ$C for 300 s. The specific contact resistance of the Ohmic contacts was measured in a transfer length method (TLM) pattern to be 1-2 $\Omega\cdot$mm at low temperatures (10 K$-$100 K). Standard UV lithography and metal evaporation were performed to deposit Ti/Au (5 nm/ 100 nm) contacts for the bottom Schottky gate. 

In addition to the front side processing steps described in the Letter, we performed extra passivation and dry etch for the devices described in this Letter. These steps were included to minimize the gap between the QW die and the substrate during the flip-chip bonding process, but are not required. For the passivation of the device, we deposited a 500-nm SiO\textsubscript{2} dielectric layer on the entire front surface after we defined the frontside contacts. To electrically access the frontside contacts after the passivation, we defined openings on the SiO\textsubscript{2} layer using dry etching. After this step, our QW sample piece was diced into smaller QW dies (3 mm x 3 mm) each of which contained a single Hall bar device, or a TACIT device, for individual flip-chip bonding. 

\subsection{Flip-chip Bonding and Underfilling}

For flip-chip bonding, Si host substrates were first prepared. Standard UV lithography and metallization steps were used to define both the electrodes and the indium bonds. For the electrodes and bonding pads, we deposited 5-nm Ti and 100-nm Au. For the indium bonds, we thermally evaporated $\sim$ 2-$\mu$m thick indium. 

For the bonding, we used a Finetch flip-chip bonder (Finetech Lambda) to align and place the QW die on the substrate. A constant force ($\sim$ 3 N) was applied on the back side of the QW die while both the die and Si host substrate were rapidly heated up to 180 $^{\circ}$C under 1 min. This causes the indium bumps to reflow and to bond with the matching electrodes on the Si substrate, directly establishing the electrical connection between the QW frontside contacts and the bonding pads on the Si host substrate. 

After the bonding, we applied a small amount of EPO-TEK 353ND underfill epoxy on one side of the flip-chip bonded sample, and waited for $\sim$ 10 mins for the epoxy to fill the gap. This waiting time is critical to minimize air bubbles on the bonded surfaces. After waiting, we moved the bonded sample to a hot plate and soft-baked it at 80 $^{\circ}$C for an hour.

\subsection{Backside Fabrication}
After the curing of the underfilli epoxy, we removed the backside GaAs ($\sim$ 520 $\mu$m thick) and the etch-stop layer to expose the back side of the QW structure. The bulk ($\sim$ 470 $\mu$m) of the backside GaAs was first mechanically removed with Al\textsubscript{2}O\textsubscript{3} films with different grit sizes; we removed the first $\sim$ 350 $\mu$m with a 30-$\mu$m grit film, then $\sim$ 100 $\mu$m with a 3-$\mu$m film, and finally the last $\sim$ 20 $\mu$m with a 0.3 $\mu$m grit film. After the mechanical lapping, the remaining GaAs layer ($\sim$50 $\mu$m) was removed with selective wet etch using a citric acid solution (with a mix ratio of 6.1:1 for 50\% premix citric acid:H\textsubscript{2}O\textsubscript{2}) for 3-4 hrs until the etch was stopped by the Al\textsubscript{0.74}Ga\textsubscript{0.26}As etch-stop layer. This Al\textsubscript{0.74}Ga\textsubscript{0.26}As etch-stop layer was in turn removed by another selective wet etch using a diluted HF solution (with a mix ratio of 1:4 for 50\% HF:H\textsubscript{2}O) for 10-15 s until the etch was stopped by the 10 nm GaAs cap layer. This left the back side of the GaAs/AlGaAs epitaxial layers ready for the final metallization step, on which the top gates were then defined using standard UV lithography and metallization. For the top gate metallization, we deposited 5-nm Ti and 100-nm Au. Before the final lithography step, the edge beads formed by the cured epoxy (which sits higher than the thinned-down GaAs/AlGaAs layers after the selective wet etches) were removed using a razor blade under a microscope for accurate patterning results.

\section{Direct Detection}

\subsection{Experimental Setup}
\begin{figure}[h]
    \centering
    \includegraphics[width=0.3\textwidth]{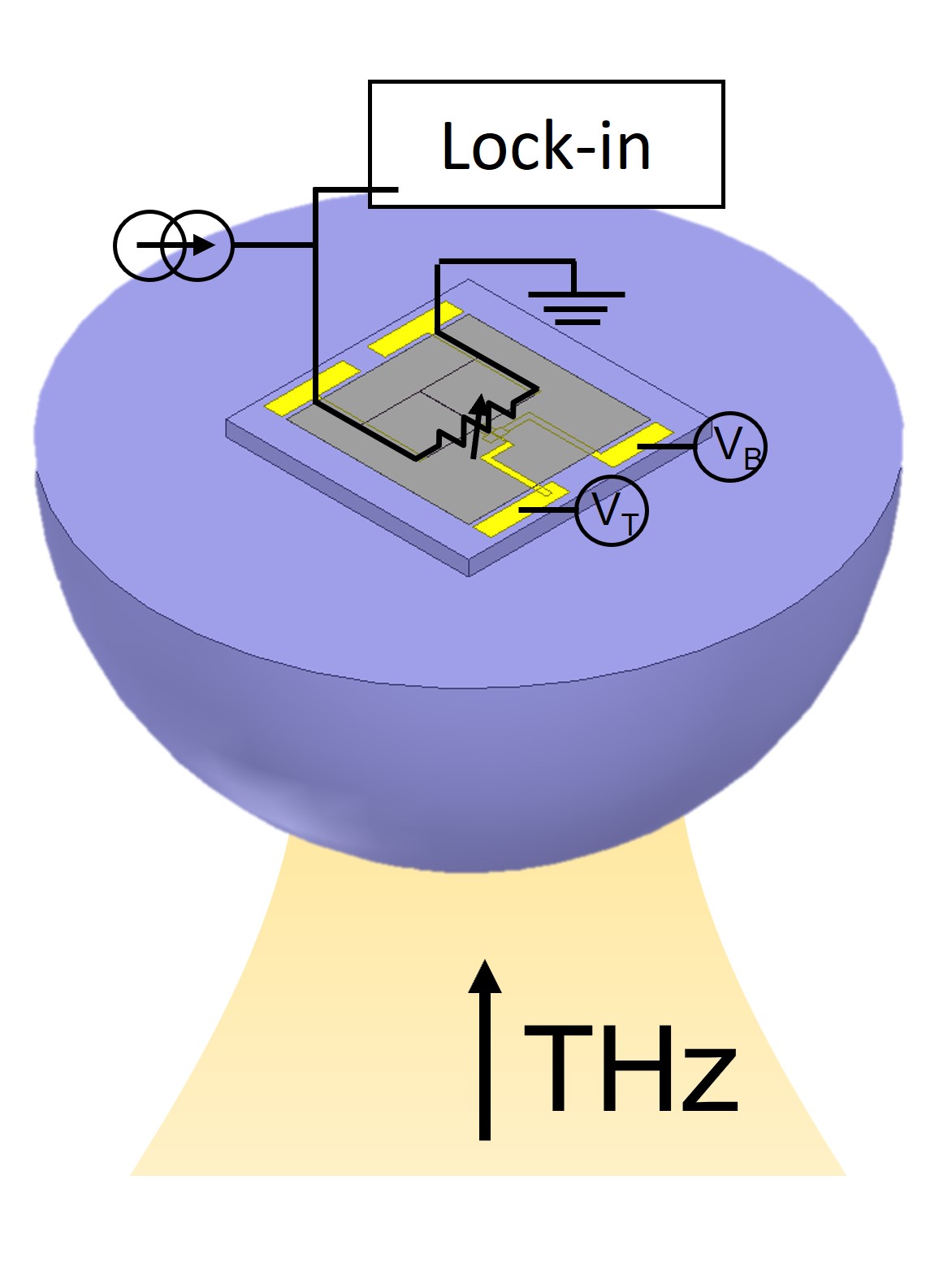}
    \caption{Schematic for the quasi-optical setup for direct detection.}
    \label{fig:SM1}
\end{figure}

For the direct detection measurements, the fabricated TACIT device was bonded to the back side of a hyper-hemispherical lens made of high-resistivity Si for quasi-optical coupling of THz radiation (Fig. \ref{fig:SM1}). The device contained a THz antenna integrated with the active region of the device, to which THz radiation was focused with the lens. The device-lens assembly was mounted on a mixer block, which in turn was thermally anchored to the cold plate inside a liquid cryogen dewar filled with liquid nitrogen and liquid helium. A bandwidth limiting filter was mounted on the 77 K thermal shield of the cryostat to filter any unwanted thermal radiation coupling to the device. For THz radiation, a quantum-cascade vertical-external-cavity surface-emitting laser (QC-VECSEL)\cite{Curwen2019, Curwen2020} was used to provide a THz signal at 3.44 THz with a nominal power on the order of 1 mW. A lock-in amplifier was used to measure the device response in the in-plane resistance of the device ($V_{DS}$ across the source-drain channel) while chopping the THz signal with a mechanical chopper.

\subsection{Model Response}

For the model response, we calculated the impedance matching efficiency between the THz antenna and the active region of the device (seen by the antenna), to which the direct detection response is expected to be linearly proportional. The expression for the impedance matching efficiency $\alpha$ is given by 
\begin{equation}
    \label{eq:alpha}
     \alpha = \frac{4R_{ant}R_{AR}}{(R_{ant}+R_{AR})^{2}+(X_{ant}+X_{AR})^{2}}
\end{equation}
where $Z_{ant}(\omega) = R_{ant} + i X_{ant}$ is the antenna impedance, and $Z_{AR}(\omega)=R_{AR}+iX_{AR}$ is the impedance of the active region of the device (seen by the antenna) at an angular frequency $\omega$.

\begin{figure}[ht]
    \centering
    \includegraphics[width=0.4\textwidth]{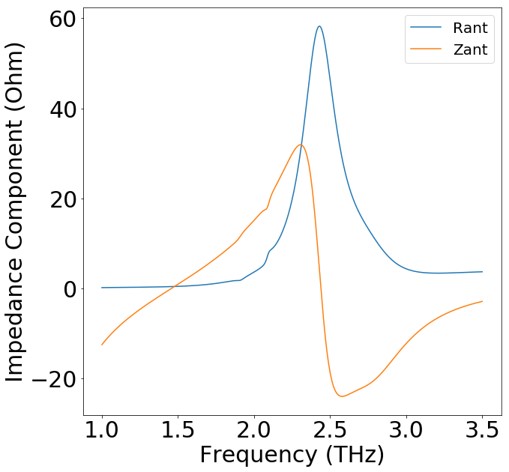}
    \caption{Calculated impedance of the antenna from ANSYS HFSS}
    \label{fig:SM2}
\end{figure}

For the antenna impedance, we used a finite-element method (FEM) software (ANSYS HFSS) to numerically estimate the impedance values for the single-slot antenna used in the prototype device. Figure \ref{fig:SM2} shows the calculated antenna impedance for the given geometry of the slot antenna. Note that the single-slot antenna used for the prototype device described in the Letter was optimized for operation near 2.5 THz and was expected to present relatively small impedance values at higher THz frequencies due to the strong frequency dependence of the antenna. From the simulation result, we estimated $Z_{ant}(3.44$ THz$) \sim 4 \Omega - i 4 \Omega$. 

For the impedance of the active region, we calculated the values based on our THz impedance model for intersubband absorption (note that this impedance is not the same as the in-plane impedance of the device).\cite{Sherwin2002,Yoo2022} The expression for the impedance is derived by considering a THz current oscillating between the top and the bottom gates and the corresponding voltage drop caused by the polarization of the 2DEG.\cite{Sherwin2002, Yoo2020, Yoo2022} The 2DEG polarization is related to 2D susceptibility of the 2DEG associated with the intersubband transitions. The general expression for the impedance of the active region $Z_{AR}(\omega)=R_{AR}+iX_{AR}$ at an angular frequency $\omega$ is given in Ref.\onlinecite{Sherwin2002} and in Ref.\onlinecite{Yoo2022}. For each component, we have
\begin{equation}
\label{eq:R_AR}
R_{AR}=\frac{e^{2}}{\epsilon_{r}^{2} \epsilon_{0}^{2} m^{*}} \frac{1}{A} \frac{2 \Gamma N_{S} f_{12} n(T_{e})}{\left(\omega^{* 2}-\omega^{2}\right)^{2}+4 \omega^{2} \Gamma^{2}}
\end{equation}
and 
\begin{equation}
\label{eq:X_AR}
X_{AR}=-\frac{1}{\epsilon_{r} \epsilon_{0} \omega} \frac{2d}{A}+\frac{1}{\epsilon^{2} \epsilon_{0}^{2} \omega A} \frac{e^{2} N_{s} f_{12} n(T)}{m^{*}} \frac{\omega^{* 2}-\omega^{2}}{\left(\omega^{* 2}-\omega^{2}\right)^{2}+4 \omega^{2} \Gamma^{2}}.
\end{equation}
where $e$ is the electron charge, $\epsilon_{r}$ is the dielectric constant, $\epsilon_{0}$ is the vacuum permittivity, $m^{*}$ is the effective mass of GaAs, $A$ is the area of the active region, $\Gamma/2\pi$ is the HWHM of the intersubband transition (between the first and the second subbands), $N_{s}$ is the total sheet charge density, $f_{12}=2m^*z_{12}^2E_{12}/\hbar^2$ is the effective oscillator strength where $z_{12}$ is the matrix element of the position operator applied on the wavefunctions for the first and second subbands, $E_{12}$ is the energy spacing between the two subbands, and $\hbar$ is the reduced Planck's constant, $n(T_{e})$ is the normalized population difference $n(T_{e})=\frac{N_{1}(T_{e})-N_{2}(T_{e})}{N_{s}}$ where $N_{1(2)}(T_{e})$ is the population of the first (second) subband at an electron temperature $T_{e}$, $\omega^{*}$ is the angular frequency for the intersubband absorption frequency, and $d$ is the distance between the center of the 2DEG and the gates.  

As shown in Eq.\ref{eq:R_AR} and in Eq.\ref{eq:X_AR}, the impedance of the active region seen by the antenna depends on both the geometric factor ($A$) and the intersubband absorption properties of a given quantum well ($\Gamma$, $N_{s}$, $f_{12}$, $n(T_{e})$, and $\omega^{*}$). The intersubband absorption properties (except for $\Gamma$) can be numerically calculated by self-consistently solving the Schr\"{o}dinger equation and the Poisson's equation for a given quantum well structure at a given temperature, charge density, and DC electric field.\cite{Helm1999,Sherwin2002,Yoo2020} For the calculation of the device impedance at 3.44 THz, we used $A \sim 25 \mu$m\textsuperscript{2} and the intersubband absorption properties (except for $\Gamma$) numerically calculated for the 40-nm quantum well at 30 K over the range of the charge density and the DC electric field used in the experiment. For $\Gamma$, we assumed a HWHM of 220 GHz at 3.44 THz, which is reasonable based on our previous experimental work with the same 40-nm quantum well structure.\cite{Williams2001}

\begin{figure}[h!]
    \centering
    \includegraphics[width=0.45\textwidth]{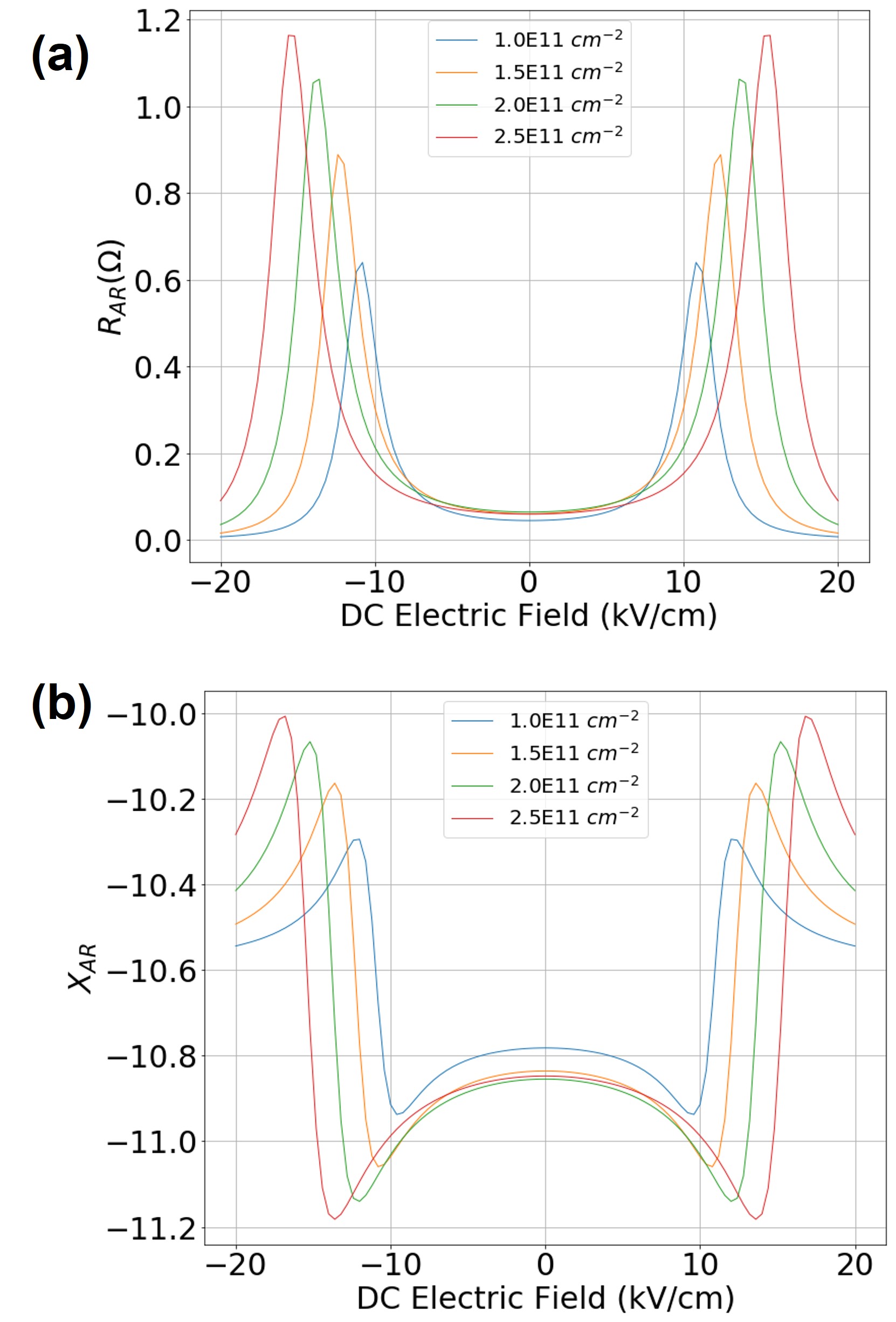}
    \caption{Calculated impedance for the active region of the device for the 40-nm quantum well at 30 K and 3.44 THz with $\Gamma/2\pi = 220$ GHz, $2d = 660 nm$, and $A = 25 \mu$m\textsuperscript{2}. (a) Real part of the impedance. (b) Imaginary part of the impedance.}
    \label{fig:SM3}
\end{figure}

Figure \ref{fig:SM3} shows the calculated impedance of the active region of the device for the 40-nm quantum well at 30 K at 3.44 THz. Each curve corresponds to the impedance values at a specific charge density indicated in the legends. The real part of the impedance is relatively small in the prototype device due to large $A$ and sub-optimal $\Gamma$ (Fig. \ref{fig:SM3}a). The imaginary part of the impedance is mostly dominated by the first term in Eq. \ref{eq:X_AR}, which is the geometric capacitance formed by the dual-gate, and approaches to the geometric capacitive reactance value of $\sim -10.6 \Omega$ (Fig. \ref{fig:SM3}b) when the incoming THz radiation matches the intersubband absorption frequency ($\omega = \omega^{*}$).

\subsection{DC Electric Field Calibration}
\begin{figure}
    \centering
    \includegraphics[width=0.45\textwidth]{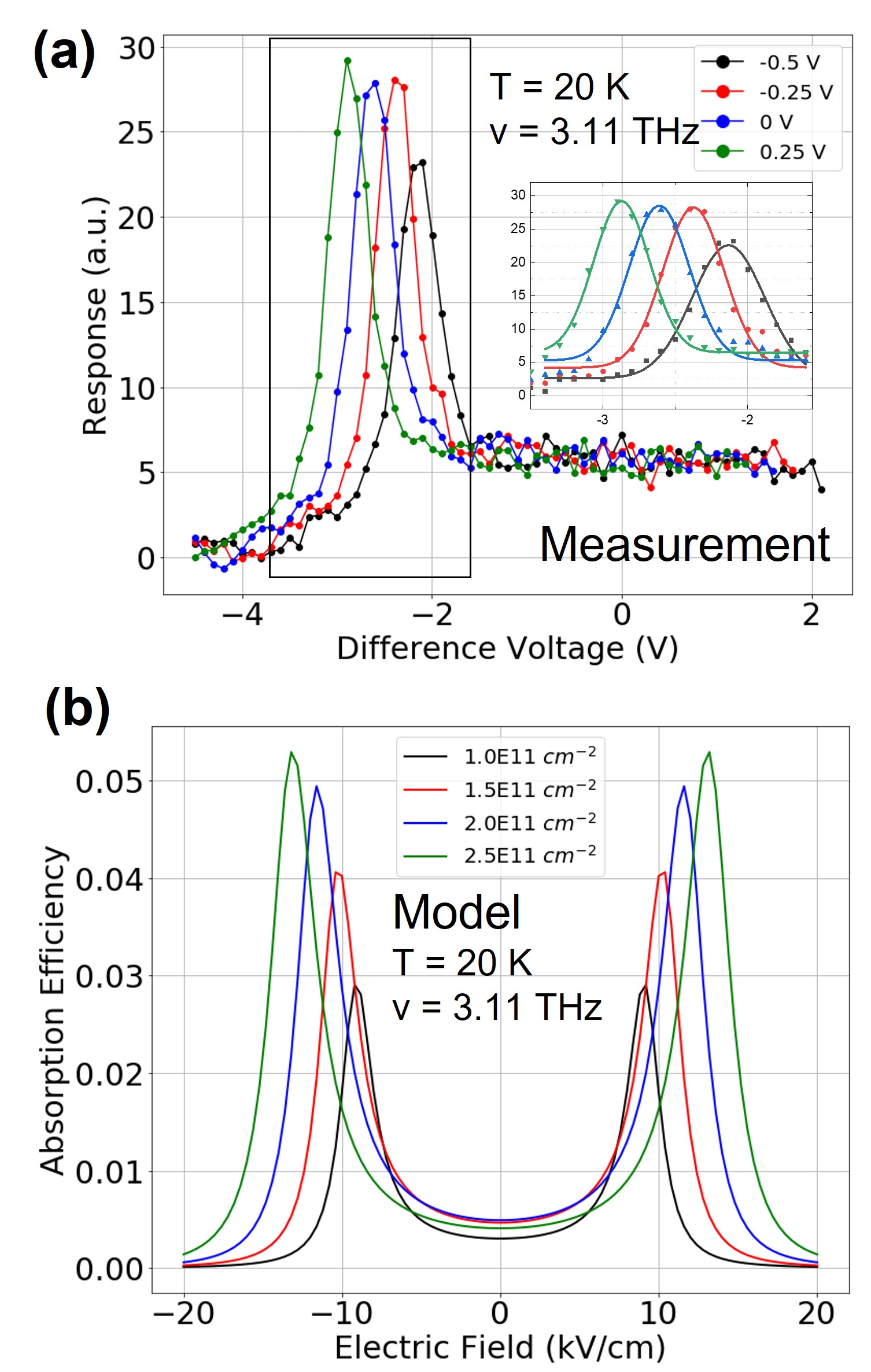}
    \caption{Shifting of intersubband absorption peaks at 3.11 THz at 20 K. (a) Experimental data. The inset shows the Gaussian fits to the peaks. (b) Model response with $\Gamma/2\pi=200 GHz$ for 3.11 THz.}
    \label{fig:SM4}
\end{figure}

\begin{figure}
    \centering
    \includegraphics[width=0.45\textwidth]{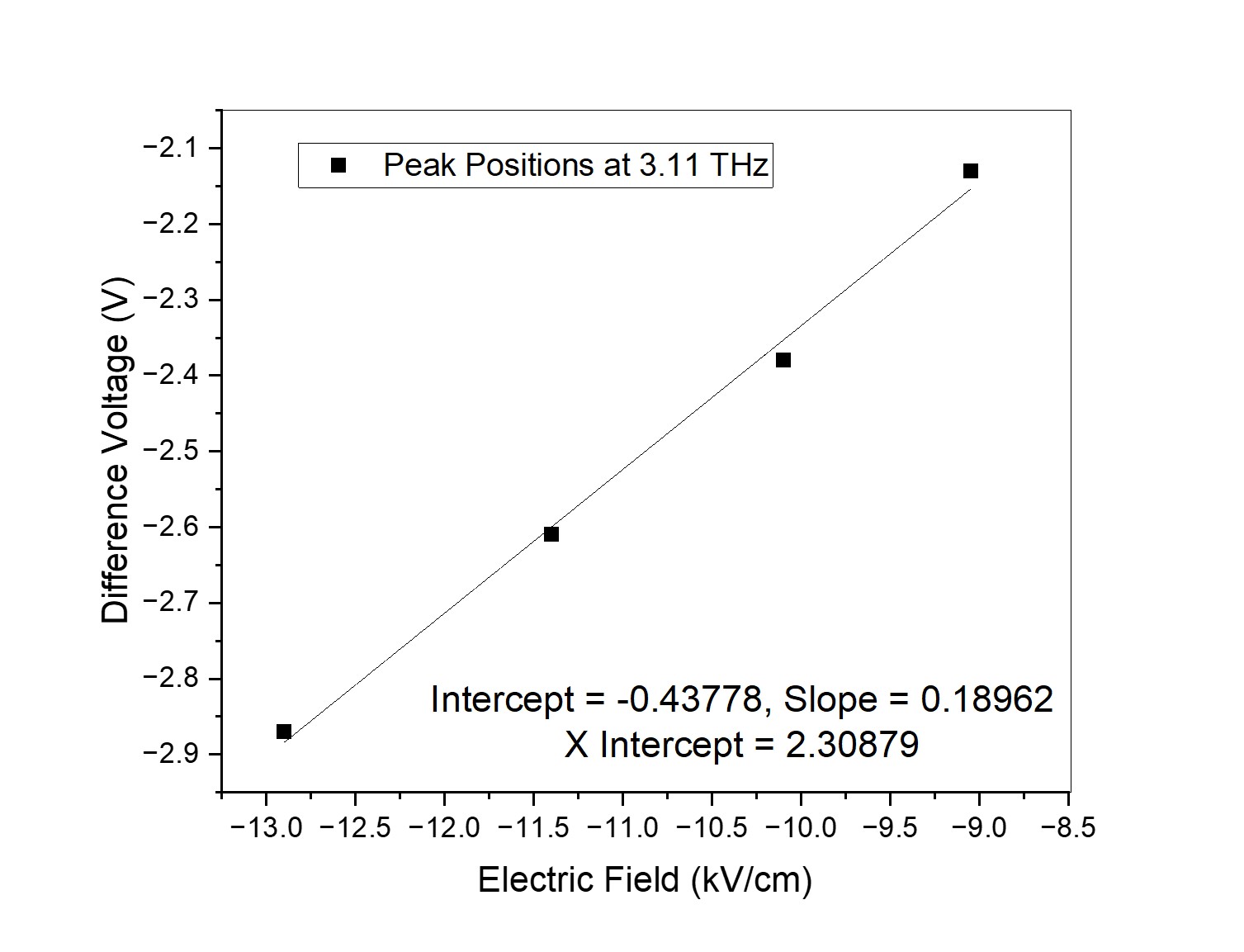}
    \caption{$E_{DC}$-to-$V_{D}$ conversion. Each point corresponds to the location of the intersubband absorption peaks at four different charge densities in the experimental data (in $V_{D}$) and in the model response (in $E_{DC}$).}
    \label{fig:SM5}
\end{figure}

To compare the direct detection results with the model response, we converted the difference voltage $V_{D}$ in the data to DC electric field $E_{DC}$ based on the calibration done at 3.11 THz. Figure \ref{fig:SM4} shows the direct detection results (Fig.\ref{fig:SM4}a) and the model response (Fig.\ref{fig:SM4}b) over four different charge densities at 3.11 THz at 20 K. Note that the responses in the data are plotted as a function of $V_{D}$, which is the experimental input parameter, while the model responses are plotted as a function of $E_{DC}$, which is the input parameter for the model. Similarly to our 3.44 THz results, we observed the outward shifting of the intersubband absorption peaks with higher response peaks over an increasing charge density. For the conversion, we recorded the locations of these peaks in the data (in $V_{D}$) by fitting the peaks with the Gaussian function (see the inset in Fig.\ref{fig:SM4}a), and compared them with the peak locations in the model (in $E_{DC}$). Assuming linear dependence of $E_{DC}$ on $V_{D}$, we acquired the following conversion relation between $E_{DC}$ (in kV/cm) and $V_{D}$ (in V): 
\begin{equation}
    V_{DC} = 0.19E_{DC}-0.44
    \label{eq:E_to_V}
\end{equation}
where the second term accounts for the built-in electric field in the quantum well (Fig. \ref{fig:SM5}). We note that the coefficient in the first term is about 3 times larger than the estimate based on a simple parallel capacitor model, in which we have $V_{DC} = 2d \times E_{DC} $ where $d$ is the distance between the metal gate and the 2DEG ($d=330 nm$ for our structure). One possible explanation is the existence of trapped charges between the metal gates and the GaAs cap layers. 

\bibliography{IBASE}